%Paper: cond-mat/9506024
%From: monthus@ipncls.in2p3.fr
%Date: Tue, 6 Jun 95 12:09:05 +0200

 \magnification\magstep 1
\vsize=24 true cm
\hsize=16 true cm
\baselineskip=14pt
\parindent=1 true cm
\parskip=0.3 true cm
\def\oppropto{\mathop{\propto}}
\def\operarrow{\mathop{\longrightarrow}}

\vglue 4 true cm

\centerline{ \bf DIFFUSION IN ONE DIMENSIONAL RANDOM MEDIUM}
\centerline{ \bf}
\centerline{ \bf AND HYPERBOLIC BROWNIAN MOTION}

\vskip 2 true cm

\centerline{   Alain Comtet \footnote{$^{1}$}{comtet@ipncls.in2p3.fr}
 and C\'ecile Monthus \footnote{$^{2}$}{monthus@ipncls.in2p3.fr} }

\vskip 1 true cm
\centerline{ {\it Division de Physique Th\'eorique} \footnote{$^{3}$}{Unit\'e
de recherche des Universit\'es Paris 11 et Paris 6 associ\'ee au CNRS} {\it ,
IPN
B\^at. 100, 91406 ORSAY, France} }
\centerline{ \bf}
\centerline{ and}
\centerline{ \bf}
\centerline{\it L.P.T.P.E., Universit\'e Paris 6, 4 Place Jussieu, 75252 PARIS,
France}

\vskip 2 true cm

\noindent { \bf Abstract}

\noindent Classical diffusion in a random medium involves an exponential
functional of
Brownian motion. This functional also appears in
the study of Brownian diffusion on a Riemann surface of constant negative
curvature.
We analyse in detail this relationship and study various distributions using
stochastic
calculus and functional integration.


\vfill

\line {IPNO/TH 95-18  \hfill}

 \eject

\leftline{\bf 1. Introduction}

\vskip 0.5 true cm

 There is a close link between one dimensional, or quasi-one dimensional,
disordered
 systems and Brownian diffusion on Riemann manifolds of constant
negative curvature. Such a
 correspondance can be traced back to the
 pioneering work of Gertsenshtein et al. $^{(1)}$ who have shown
 that statistical properties
of reflection and transmission coefficients of waveguides with random
inhomogeneities
are directly related to some random walk on the Lobachevsky plane.
There has been a renewed interest in this approach through the study of
mesoscopic
systems. The description of quasi-one dimensional mesoscopic wires involves
a Fokker-Planck equation giving the probability distribution of the N
eigenvalues of
the transmission matrix $^{(2)}$. Recently it has been shown that this equation
can be
interpreted as the diffusion equation on a Riemannian symmetric space $^{(3)}$.

The purpose of this work is to show how the one dimensional classical diffusion
of a particle in a quenched random potential $U(x)$
 that is itself a Brownian motion, possibly with some constant drift,
 is directly related to Brownian motion on the hyperbolic plane.
Since the latter is the archetype of chaotic systems $^{(4)}$, our work forms a
bridge between
disordered and chaotic systems.

\vskip 1 true cm

\leftline{ \bf 2. Fundamental random variable for one dimensional classical
diffusion}
\leftline{ \bf \quad in a quenched random potential $U(x)$}

\vskip 0.5 true cm

A large amount of work has been devoted to random random walks, defined on a
lattice
by the following Master equation
$$
P_n (t+1) = \alpha_{n-1} P_{n-1} (t) + \beta_{n+1} P_{n+1}
\eqno (2.1)
$$
where $\alpha_{n}$ is the random quenched transition rate from site $n$ to site
$(n+1)$, and
$\beta_{n} \equiv 1- \alpha_{n}$ is the random quenched transition rate from
site $n$ to site $(n-1)$.
 From a physical point of view it is convenient to introduce a corresponding
random
potential $U(n)$ on each site $n$
 and to write the ratio
of the two transition rates $\alpha_{n}$ and $\beta_{n}$ as an Arrh\'enius
factor
$$
\sigma_n \equiv {{\beta_{n}} \over {\alpha_{n}}} =
 { {e^{- \beta \left[ U(n-1) -U(n) \right]}}
\over {e^{- \beta \left[ U(n+1) -U(n) \right]}}}
= {e^{ \beta \left[ U(n+1) -U(n-1) \right]} }
\eqno (2.2)
$$
The study of different physical quantities related to this random random walk
$^{(5)}$ involves systematically
random variables of the following form
$$
Z(a,b) = \sum_{n=a}^{b} \prod_{k=a}^{n} \sigma_k = \sigma_a + \sigma_a
\sigma_{a+1}
+ \cdots + \sigma_a \sigma_{a+1} \ldots \sigma_b
\eqno (2.3)
$$
Their fundamental property is to satisfy the linear random coefficient
 recurrence relation
$$
Z(a,b) = \sigma_a \bigg[ 1+Z(a+1,b) \bigg]
\eqno (2.4)
$$
This discrete multiplicative stochastic process also appears for the Ising
chain in a
random magnetic field $^{(6)}$.

Let us now consider the continuous model of classical diffusion on the line
defined by
 the Fokker-Planck equation for the probability density $P(x,t \  \vert x_0,
0)$
$$
{{\partial P} \over {\partial t }} = {1 \over 2} {\partial \over {\partial x}}
\left( {\partial P\over {\partial x}} - \beta F(x) P \right)
\eqno (2.5)
$$
where $\{F(x)\}$ is a quenched random force.
In this continuous limit the discrete random variable $Z(a,b)$ defined in (2.3)
becomes
the following exponential functional of the random potential $U(x)=-\int^x F(y)
dy$
$$
\tau(a,b) = \int_a^b dx \ e^{ \displaystyle \beta \left[U(x)-U(a) \right ]}
 = \int_a^b dx \ e^{ \displaystyle - \beta \int_a^x F(y) dy}
\eqno (2.6)
$$
The evolution of this functional is governed by the
stochastic differential equation
$$
{{\partial \tau } \over {\partial a }}(a,b) = \beta F(a) \tau(a,b) -1
\eqno (2.7)
$$
which replaces the random coefficient recurrence relation (2.4) satisfied by
$Z(a,b)$.
Note that the stochastic term $F(a)$ appears multiplicatively, so that the
fluctuations of the random force are coupled to the values taken by the random
process
$\tau(a,b)$.

 Let us now make clear how the functional $\tau(a,b)$ arises in some
physical quantities associated with the classical diffusion of a particle in
the
quenched random environment.

\noindent $\bullet$ If the random force has a positive mean $<F(x)> \equiv F_0
>0$,
the probability distribution of the random functional
$$
\tau_{\infty} \equiv \lim_{(b-a) \to \infty} \tau(a,b)
\eqno (2.8)
$$
determines the large time anomalous behavior of the Brownian particle position
$^{(7)}$. In particular, the velocity defined for each sample as
$$
V = \lim_{t \to \infty} {d \over dt} \int_{- \infty}^{+ \infty}
 dx \ x P(x,t \  \vert x_0, 0)
\eqno (2.9)
$$
is a self-averaging quantity inversely proportional to the first moment of
 $\tau_{\infty}$ $^{(7)}$
$$
V= {1 \over {2 <\tau_{\infty} >}}
\eqno (2.10)
$$
When the quenched random force is distributed with the Gaussian measure
$$
{\cal D} F(x) \ e^{ \displaystyle -{1 \over {2 \sigma} } \int dx \
[F(x)-F_0]^2}
\eqno (2.11)
$$
the probability distribution ${\cal P} _{\infty} (\tau)$ of the functional
$\tau_{\infty}$ reads $^{(7)}$
$$
{\cal P} _{\infty} (\tau) = {\alpha \over \Gamma(\mu)} \left( {1 \over \alpha
\tau}
\right)^{1+\mu} \ e^{ \displaystyle - {1 \over \alpha \tau}}
\ \ \oppropto_{ \tau \to \infty} \ \ {1 \over \tau^{1+\mu}}
\eqno (2.12)
$$
where
$\mu = {2 F_0 \over {\beta \sigma}} >0$ is a dimensionless parameter and
 $ { \alpha} ={ {\sigma \beta^2} \over 2}$.
The algebraic decay at large $\tau$ accounts for the dynamical phase
transitions
which occur in this model $^{(7)}$. In particular,
equation (2.10) implies that the value $\mu = 1$ separates a phase of vanishing
velocity $V=0$ for $0<\mu<1$, and a phase of finite velocity $V>0$ for $\mu
>1$.

\noindent $\bullet$ The functional $\tau(a,b)$
also arises in the study of transport properties of finite size
disordered samples. The stationary current $J_N$ which goes through a
disordered
sample of
length $N$ with fixed concentrations $P_0$ and $P_N$ at the boundary can be
written in terms
of the exponential functional $\tau_N \equiv \tau(0,N)$ as $^{(8)}$ $^{(9)}$
$$
J_N = {1 \over 2} \left[ {P_0 \over \tau_N } -
P_N { {{\partial \ln \tau_N  } \over {\partial N  }} }  \right]
\eqno (2.13)
$$
When the end $x=N$ is a trap described by the boundary condition $P_N=0$, the
flux
$J_N$ is simply a random variable inversely proportional to $\tau_N$.
The probability distribution of $\tau_N$ has been studied for the case of zero
mean force $F_0=0$ $^{(8)}$, and for the general case with arbitrary
mean force $^{(9)}$ by different methods.

It is also interesting to point out that a lot of mathematical work
has recently been devoted to the functional $\tau_N$ in relation with finance
$^{(10)}$ $^{(11)}$ $^{(12)}$. \hfill \break
In $^{(11)}$, Yor pointed out the relation between the functional $\tau_N$
for the particular case $\mu={1\over 2}$, and free Brownian motion on the
hyperbolic plane.
In the following, we first rederive this correspondance,
and then generalize it to arbitrary
$\mu$, using some external drift on the hyperbolic plane.

\vskip 1 true cm

\leftline{ \bf  3. Relation with hyperbolic Brownian motion}

\vskip 0.5 true cm

The upper half-plane $\{\  (x,y) \ ,  \ y>0 \}$ endowed with the metric
$$
ds^2 = { {dx^2 + dy^2} \over y^2}
\eqno (3.1)
$$
defines a two-dimensional Riemann manifold of constant negative Gaussian
curvature $R= -1$.
The surface element $dS$ and the Laplace operator $\Delta$ are covariantly
defined as
$$
dS = { {\ dx \ dy} \over y^2}   \ \ \ \ \ \  \hbox{and} \ \ \ \ \ \
\Delta={y^2 }
\left( {{\partial^2} \over {\partial x ^2}} + {{\partial^2} \over {\partial y
^2}} \right)
\eqno (3.2)
$$
Free Brownian motion on this manifold is defined by the
 diffusion equation for the Green's function $G_t (x,y)$
$$
{ {\partial G} \over {\partial t}} = D \Delta  G = D \ {y^2 }
\left( {{\partial^2} \over {\partial x ^2}} + {{\partial^2} \over {\partial y
^2}} \right) G
\eqno (3.3)
$$
It is convenient to choose the initial condition at the point $(x=0, y=1)$
$$
G_t (x,y) \operarrow_{t \to 0^+} \delta(x) \ \delta(y-1)
\eqno (3.4)
$$
The normalization of the Green function $G_t (x,y)$
 then reads for any time t
$$
1=\int dS  \ G_t (x,y) = \int_{-\infty}^{+\infty} dx \ \int_0^{+ \infty} dy
\ {1 \over y^2}\ G_t (x,y)
\eqno (3.5)
$$
 Consider the probability density $P_t(x,y)$
$$
P_t(x,y) = {1 \over y^2}\  G_t (x,y)
\eqno (3.6)
$$
normalized with respect to the flat measure $dx \ dy $
$$
1 = \int_{-\infty}^{+\infty} dx \ \int_0^{+ \infty} dy \  P_t (x,y)
\eqno (3.7)
$$
This probability density satisfies the Fokker-Planck equation
$$
{ {\partial P} \over {\partial t}}  = D  \left( {{\partial^2} \over {\partial x
^2}} +
{{\partial^2} \over {\partial y ^2}} \right) \ {y^2 }\  P
\eqno (3.8)
$$
and the initial condition at $t=0$
$$
P_t (x,y) \operarrow_{t \to 0^+} \delta(x) \delta(y-1)
\eqno (3.9)
$$
We now introduce two independant Gaussian white noises
  $\eta_1(t)$ and $\eta_2(t)$
and write the stochastic differential equations for the process $\{x(t),
y(t)\}$
corresponding to the Fokker-Planck equation (3.8)
following respectively the It\^o or Stratonovich convention $^{(13)}$
$$  \hbox{ It\^o}   \left\{\matrix{
\displaystyle {dx\over dt}={ \sqrt {2D}} \ {y } \ \eta_1(t)\hfill\cr
\noalign{\medskip}
\displaystyle {dy\over dt}={ \sqrt {2D}} \ {y} \ \eta_2(t)\hfill\cr
}\right\}
\eqno (3.10)
$$
$$ \hbox{Stratonovich} \left\{\matrix{
\displaystyle{dx\over dt}={ \sqrt {2D}}\  {y } \ \eta_1(t)\hfill\cr
\noalign{\medskip}
\displaystyle {dy\over dt}=  - {D }\  y +{ \sqrt {2D}} \ {y } \
\eta_2(t)\hfill\cr
}\right\}
\eqno (3.11)
$$
 Integration of these two coupled differential equations gives
$$\left\{\matrix{
\displaystyle x(t) = { \sqrt {2D}} \int_0^t dv  \ \eta_1(v) \
e^{ \displaystyle - {D }v + {{ \sqrt {2D}} } \int_0^v \eta_2(s) ds }\cr
\hbox{} \cr
\displaystyle y(t)=   \ e^{ \displaystyle - {D }t + {{ \sqrt {2D}}  }
\int_0^t \eta_2(s) ds } \hfill \cr
}\right\}
\eqno (3.12)
$$
The process $y(t)$ is therefore simply the exponential of some linear Brownian
motion
with negative drift. Let us now study more precisely the process $\{x(t)\}$.
Since the white noise $\eta_1$ appears linearly in $x(t)$, all the odd moments
therefore
vanish
$$
< x^{2n+1}(t)> _{\eta_1} = 0
\eqno (3.13)
$$
and all the even moments can be rewritten as averages over $\eta_2$ only. In
particular the second
moment reads
$$
< x^2(t)> _{\eta_1, \eta_2} = 2D \
< \ \int_0^t dv \ e^{ \displaystyle - {2 D }  v +
 2{{ \sqrt {2D}}  } \int_0^v \eta_2(s) ds } \ > _{\eta_2}
\eqno (3.14)
$$
If we set
$$
\beta \int_0^v F(u) du = {2 D}  v - 2{  \sqrt {2D} } \int_0^v \eta_2(s) ds
\eqno (3.15)
$$
the expression between brackets on the right hand side of Eq (3.14) is nothing
but the functional
$$
\tau_t = \int_0^t \ dv \ e^{ \displaystyle -\beta \int_0^v F(u) du}
\eqno (3.16)
$$
which is encountered in the study of classical diffusion in a quenched force
$\{F\}$
distributed as a Gaussian white noise
$$
{\cal D} \  F (x) \  e^{ \displaystyle -{1 \over 2 \sigma} \int_{- \infty}^{+
\infty}
[F(x) - F_0]^2 dx}
\eqno (3.17)
$$
with parameters $F_0 = \displaystyle  \ {2D \over \beta } $ and
$\sigma = 8 \ \displaystyle {D \over \beta^2 }  $.
More generally, all even moments of $x(t)$ are proportional to moments of
the functional $\tau_t$
$$
< x^{2n}(t)> _{\eta_1, \eta_2} = (2D)^n \ < \tau^n_t > _{F} \ \ {\cal N}(n)
\eqno (3.18)
$$
 ${\cal N}(n)$ being the combinatorial factor coming from Wick's theorem
that counts the number of ways to pair the $2n$ fonctions $\eta_1$.
In particular, ${\cal N}(n)$ is equal to the moment of order $2n$
of a suitable Gaussian random variable $\xi$ with variance unity
$$
{\cal N}(n) =\int_{- \infty}^{+ \infty} {{d \xi} \over {\sqrt {2\pi}}} \
{\xi}^{2n} \
e^{ \displaystyle   -{\xi^2 \over 2}} =2^n \ {{\Gamma (n+{1 \over 2})}
\over {\Gamma ({1 \over 2})}}= \prod_{k=1}^n (2k-1)
\eqno (3.19)
$$
These identities between moments allow us to rewrite the process $x(t)$ in the
very
simple form suggested by Yor $^{(11)}$
$$
x(t) ={ \sqrt {2D}} \int_0^{\displaystyle \tau_t} du \ \eta(u)
\eqno (3.20)
$$
where $\eta$ is a new independent Gaussian white noise.
The process $x(t)$ is therefore identical in law to a linear Brownian motion
whose effective time is the random variable $\tau_t$.
This result can also be obtained through a functional integration method using
an appropriate change of time in path-integral (see Appendix).

To sum up, the free Brownian motion $\{x(t), y(t)\}$ on the hyperbolic plane
can be rewritten in terms of the two independant white noises of measure
$$
{\cal D} \  F (x) \  e^{ \displaystyle -{1 \over 2 \sigma} \int_{- \infty}^{+
\infty}
[F(x) - F_0]^2 dx}\ \ \
{\cal D} \eta(t)\
\ e^{ \displaystyle - {1 \over 2} \int dt \ \eta^2(t) }
\eqno (3.21)
$$
The process $y(t)$ is simply the exponential of the Brownian motion with drift
\hfill \break
 $U(t) =- \int_0^t F(u) du$
$$
y(t) =  \ e^{ \displaystyle - {\beta \over 2} \int_0^t F(u) du} = \  e^{
\displaystyle
{\beta \over 2}  U(t)}
\eqno (3.22)
$$
The process $x(t)$ can be viewed as a linear Brownian motion
$$
x(t) ={ \sqrt {2D}}\  \int_0^{\displaystyle \tau_t} du \ \eta(u)
\eqno (3.23)
$$
with an effective time $\tau_t$ that is itself a random process
depending on $y(t)$
$$
\tau_t = \int_0^t dv \ e^{ \displaystyle \beta U(v)} =
 \int_0^t\  dv \ {y^2(v) }
\eqno (3.24)
$$
Note that the dimensionless parameter $\mu \equiv {2 F_0 \over {\beta \sigma}}$
which characterizes the different phases of anomalous diffusion
 is  ${1 \over 2}$ for the free
hyperbolic Brownian motion. This value represents the natural drift
 induced by the curvature of the Poincar\'e half-plane.
 It is, however, easy to generalize the previous analysis to arbitrary $\mu$
with the introduction of some external constant drift $m$ in direction $y$
$$
\mu ={ 1 \over 2} +m
\eqno (3.25)
$$
The corresponding stochastic differential equations then read
$$  \hbox{ It\^o}   \left\{\matrix{
\displaystyle {dx\over dt}={ \sqrt {2D}} \ {y } \ \eta_1(t)\hfill\cr
\noalign{\medskip}
\displaystyle {dy\over dt}= - 2D \ m \  y +{ \sqrt {2D}} \ {y} \
\eta_2(t)\hfill\cr
}\right\}
\eqno (3.26)
$$
$$ \hbox{Stratonovich} \left\{\matrix{
\displaystyle{dx\over dt}={ \sqrt {2D}}\  {y } \ \eta_1(t)\hfill\cr
\noalign{\medskip}
\displaystyle {dy\over dt}=  - 2D \mu \  y +{ \sqrt {2D}} \ {y } \
\eta_2(t)\hfill\cr
}\right\}
\eqno (3.27)
$$

 For any $\mu$, there is therefore a direct correspondance through Eqs
(3.22-3.23)
between
the joint stochastic process characterizing the one dimensional diffusion
$\{$ random potential $U(t)$, exponential
functional $\tau_t \} $ and the Brownian motion $\{x(t), y(t)\}$ on the
hyperbolic plane
with possibly some external constant drift $m$ along direction $y$.
We now consider some consequences of this correspondance.

\vskip 1 true cm

\leftline{ \bf 4. Marginal laws of the processes $\tau_t$ \ , $x(t)$ and
$y(t)$}

\vskip 0.5 true cm

The marginal law $Y_t(y)$ of the process $y(t)$ reads according to eq (3.22)
$$
Y_t(y) = \int_{U(0)=0}{\cal D}U(s)\ e^{- \displaystyle{{1\over 2\sigma}\int_0^t
\left({dU\over ds}+F_0\right)^2 ds }} \
\delta \left(y-  a \ e^{ \displaystyle {\beta \over 2} U(t)} \right)
\eqno(4.1)
$$
We therefore get after some algebra the following log-normal distribution
$$
Y_t(y) = { 1 \over y \sqrt{ 4 \pi D t}} \ e^{\displaystyle -{1 \over { 4D t}}
\big[\ln \left({y }\right) + 2 \mu {{D } t} \big]^2 }
\eqno(4.2)
$$
where $\mu= {2F_0 \over \beta \sigma}$ and $D =
{\beta^2 \sigma \over 8}$.
In the case of free Brownian motion $(\mu={1 \over 2})$, this marginal law
tends to a $\delta$ distribution in the limit $t \to \infty$
$$
Y_{\infty}(y) = \delta(y)
\eqno(4.3)
$$
The Brownian particle is therefore attracted to the $y=0$ axis as a result of
the curvature of
the hyperbolic plane. Note that this axis represents infinity on this plane.
This limit law remains unchanged as long as $\mu \equiv \left( {1 \over 2}+m
\right) >0$.
However, when the constant external drift $m$ along $y$ is negative enough
to overcome the natural drift of the hyperbolic plane
 ($m < -{1 \over 2}$), there is no equilibrium distribution
 for the process $y(t)$.
In fact for $\mu>0$, the existence of this stationary distribution for the
process $y(t)$
governs the existence of a stationary distribution for the process $x(t)$ that
we now
construct.

The process $x(t)$ is a Brownian motion of effective time
 $\tau_t$ which is a functional of the process y(t)
$$
x(t) ={ \sqrt {2D}}\  \int_0^{\displaystyle \tau_t} du \ \eta(u)
\ \ \ \ \hbox{with} \ \ \ \
\tau_t = \int_0^t dv \ e^{ \displaystyle \beta U(v)} = \int_0^t\  dv \ {y^2(v)
}
\eqno(4.4)
$$
The statistical independence of $\eta(t)$ and $\tau_t$
allows us to write the marginal law
$X_t(x)$ of the process $x(t)$ in terms of the probability distribution
$\psi_t(\tau)$
of the functional $\tau_t$
$$
X_t(x)
= \int_0^{\infty} d \tau \ \psi_t(\tau) \ {1 \over \sqrt{4 \pi D \tau }} \
e^{\displaystyle - {x^2 \over {4D \tau}}}
\eqno(4.5)
$$
In a previous work $^{(9)}$, we showed that $\psi_t(\tau)$
satisfies a Fokker-Planck equation,
 associated with the stochastic differential equation (2.7)
which in the Stratonovich prescription reads
$$
{\partial \psi_t (\tau) \over \partial t} = {\partial \over \partial \tau}
\bigg[ \alpha \tau^2 {\partial \psi_N (\tau) \over \partial \tau} +
 \bigg( (\mu + 1)\alpha \tau - 1 \bigg) \psi_N (\tau) \bigg]
\eqno(4.6)
$$
It has to be supplemented by the initial condition $\psi_{t=0} (\tau) =
\delta(\tau)$.
We recall the solution found in terms of an expansion in the Fokker-Planck
eigenvectors basis
$$
\psi_t (\tau)=\sum_{0\leq n<{\mu\over 2}}e^{-\alpha tn(\mu-n)}
{(-1)^n(\mu-2n)\over \Gamma(1+\mu-n)}
\left( {1\over \alpha\tau} \right)^{1+\mu-n}
L_n^{\mu-2n} \left({1\over \alpha\tau} \right)e^{- \displaystyle {1 \over
\alpha\tau}}
$$
\hbox{}
$$
+{\alpha\over 4\pi^2}\int_0^{\infty} \! ds\ e^{-{\alpha t\over 4}(\mu^2+s^2)}
s\sinh \pi s \left\vert \Gamma\left(-{\mu\over 2}+i {s\over 2}\right)
\right\vert^2
\left({1\over \alpha\tau}\right)^{1+\mu\over 2} \! W_{{1+\mu\over 2},i{s\over
2}}
\left({1\over \alpha\tau}\right)e^{-\displaystyle {1\over 2\alpha\tau}}
\eqno(4.7)
$$
This leads after integration with the Gaussian kernel (4.5)
and the transposition $\alpha = 4 D$ to an expansion presenting the same time
relaxation spectrum
$$
X_t (x)=  \sum_{0\leq n<{\mu\over 2}} e^{\displaystyle - {4D } n(\mu-n) t}
{(-1)^n(\mu-2n)\over n!\Gamma(1+\mu-n)} {{\Gamma\left(\mu +{1 \over
2}-n\right)} \over
{\Gamma\left({1 \over 2}-n\right)}}\
\left( {1 \over {1+x^2}} \right)^{\mu+{1 \over 2}}
$$
\hbox{}
$$
\times\  F\left(-n,\ n-\mu,\  {1 \over 2}; \ -{x^2 }\right)
$$
\hbox{}
$$
+{1\over 4\pi^2  {\sqrt\pi}}\int_0^{\infty} \! ds\
e^{\displaystyle -{ {D}}(\mu^2+s^2)t}
s\sinh \pi s \left\vert \Gamma\left(-{\mu\over 2}+i {s\over 2}\right)
\right\vert^2
\left\vert \Gamma\left({{\mu +1}\over 2}+i {s\over 2}\right) \right\vert^2
$$
\hbox{}
$$
\times \  F\left({{\mu +1}\over 2}+i {s\over 2},\ {{\mu +1}\over 2}-i
{s\over 2},\  {1 \over 2};\  -{x^2 } \right)
\eqno(4.8)
$$
where $F(a,b,c;z)$ denotes the hypergeometric function of parameters $(a,b,c)$.

For $\mu>0$, there exists an equilibrium distribution $X_{\infty}(x)$
$$
X_{\infty}(x) = \displaystyle {1 \over {  \sqrt \pi}} \
 { {\Gamma(\mu+{1 \over 2})} \over { \Gamma(\mu)} } \
 \left( {1 \over {1+x^2}}\right)^{\mu+{1 \over 2}}
\eqno(4.9)
$$
The existence of this limit law for hyperbolic Brownian motion reveals
a "localization" phenomenon in direction $x$. This effect comes from the
attraction
towards the axis $y=0$.
Note that for the free case $(\mu={1 \over 2})$
the asymptotic marginal law $X_{\infty}(x)$ is simply a Lorentzian
$$
X_{\infty}(x) = {1 \over \pi} {1 \over {1 +x^2}}
\eqno(4.10)
$$

In summary, there exist equilibrium distributions for $Y_{\infty}$ and
$X_{\infty}$
as long as \hfill \break $\mu= \left({1 \over 2} + m \right) >0$
$$\left\{
\matrix{
&Y_{\infty}(y)= \delta(y) \hfill \cr
&\hbox{} \cr
&X_{\infty}(x) = \displaystyle {1 \over { \sqrt \pi}} \
 { {\Gamma(\mu+{1 \over 2})} \over { \Gamma(\mu)} } \
 \left( {1 \over {1+x^2}}\right)^{\mu+{1 \over 2}} \cr
}\right\}
\eqno(4.11)
$$
The joint law of the processes $\{x(t), y(t)\}$ therefore presents the
following factorized form in the limit
 $t \to \infty$
$$
P_{\infty} (x,y) =  X_{\infty}(x) \delta(y)
\eqno(4.12)
$$
However, as long as time $t$ is finite, the two processes $x(t)$ and $y(t)$
remain coupled. The study of their joint law is then needed to get a complete
description
of hyperbolic Brownian motion.

\vskip 1 true cm

\leftline{ \bf 5. Joint laws of the processes $\{ \tau_t, U(t) \}$ and $\{x(t),
y(t)\}$ }

\vskip 0.5 true cm

Path-integral methods allow us to obtain very simply the joint law
 $\psi_t(\tau \parallel u)$ of the random variables
$$
\tau _t  = \int_0^t d x \  e^{\displaystyle  \beta U(x)}
\ \ \ \ \ \hbox{and} \ \ \ \ \ U(t)
$$
Let us generalize our previous approach $^{(9)}$. The $\tau$-Laplace transform
of the joint law
$$
E \left( e^{ \displaystyle -p\tau_t} \parallel u \right)
\equiv \int_0^{\infty} d \tau \
e^{\displaystyle -p \tau} \  \psi_t(\tau \parallel u)
\eqno(5.1)
$$
can be written as a path-integral over the random potential
$$
E \left( e^{ \displaystyle -p\tau_t} \parallel u \right)
=\int_{U(0)=0}^{U(t)=u} {\cal D}U(x)\ e^{- \displaystyle{{1\over
2\sigma}\int_0^t
\left({dU\over dx}+F_0\right)^2 dx
-p\int_0^t dx \  e^{\beta U(x)}}}
$$
\hbox{}
$$
=e^{- \displaystyle{F_0^2t\over2\sigma}} e^{- \displaystyle{{F_0\over\sigma}\
u}}
\int_{U(0)=0}^{U(t)=u} {\cal D}U(x)\ \
e^{ \displaystyle - {1\over 2\sigma}\int_0^t
\left({dU\over dx}\right)^2 dx -p\int_0^t dx e^{\beta U(x)} }
\eqno(5.2)
$$
The remaining path-integral is simply the Euclidean quantum mechanics Green's
function
$<u\vert e^{-t H}\vert0>$ associated with the Liouville Hamiltonian
$$
H= -{\sigma \over 2}\  {d^2 \over du^2} + p \  e^{\displaystyle \beta u}
\eqno(5.3)
$$
We therefore get
$$
E \left( e^{ \displaystyle -p\tau_t} \parallel u \right)
= \ e^{ \displaystyle - {F_0^2t\over2\sigma}} \ \ e^{-\displaystyle
{{F_0\over\sigma}\ u}}\quad <u\vert e^{-t H}\vert 0>
\eqno(5.4)
$$
The expansion of the Green's function $<u\vert e^{-tH}\vert 0>$ in the basis of
eigenfunctions $\psi_k(u)$ of the Hamiltonian $H$
$$
\psi_k(u)=2\sqrt{{\beta k\over\alpha\pi}\sinh{2k\pi\over\sqrt\alpha}} \
K_{2ik/\sqrt\alpha} \left(2\sqrt{{p\over\alpha}}e^{\beta u/2}\right)
\eqno(5.5)
$$
gives
$$
<u\vert e^{-tH}\vert0>=\int_{-\infty}^{+\infty} { dk \over{ 2 \pi}} \ \
\psi_k(u) \ \psi^\ast_k(0)  \ e^{-k^2t}
\eqno(5.6)
$$
We finally obtain that the $\tau$-Laplace transform of
the joint law $\phi_t(\tau \parallel y)$ of the random variables
$$
\tau _t  = \int_0^t d x \  e^{\displaystyle  \beta U(x)}
\ \ \ \ \ \hbox{and} \ \ \ \ \ y(t) = e^{\displaystyle { \beta \over 2} U(t)}
$$
reads after some changes of variables
$$
E \left( e^{ \displaystyle -p\tau_t } \parallel y \right)
\equiv \int_0^{\infty} d \tau \
e^{\displaystyle -p \tau} \  \phi_t(\tau \parallel y)
$$
$$
={ e^{\displaystyle -{\alpha t\over 4}\mu^2} \over {\pi ^2}} \
{1\over y^{1+\mu}}\!
\int_{-\infty}^{+\infty}dq\  e^{\displaystyle -{\alpha t\over 4}q^2} \
q\sinh \pi q \ K_{iq}\left(2y \sqrt{p\over\alpha}\right)\
K_{iq}\left(2 \sqrt{p\over\alpha}\right)
\eqno(5.7)
$$

Let us now compute the  $x^2$-Laplace transform of the joint law
 $Q_t(x,y)$ of the hyperbolic Brownian motion starting from the series
of moments of $x^2(t)$, with $y(t)$ being fixed as $y$
$$
E \left( e^{\displaystyle -q x^2(t)} \parallel y \right) \equiv \int_{-
\infty}^{\infty}
dx \ e^{\displaystyle -q x^2} \  Q_t(x,y)
$$
$$
= \sum_{n=0}^{\infty} { {  (-q)^n }\over {n!} }   \  E \left( x^{2n}(t)
\parallel
y(t)=y \right)
\eqno(5.8)
$$
Eq (3.23) allows us to write the following relation between
moments of $x^2(t)$ and moments of $\tau_t$ when $y(t)$ is fixed as $y$
$$
 E \left( x^{2n}(t) \parallel y \right) = (2D)^n \ {\cal N}(n)
 E \left( \tau_t^{n}(t) \parallel y \right) \
\eqno(5.9)
$$
with the combinatorial factor ${\cal N}(n)$ introduced in (3.18-3.19).
Using the integral representation
$$
{\cal N}(n) =\int_{- \infty}^{+ \infty} {{d \xi} \over {\sqrt {2\pi}}} \
{\xi}^{2n}\
 e^{\displaystyle -{\xi^2 \over 2}}
\eqno(5.10)
$$
we can resum the series of moments of $\tau_t$ under the integral
$$
 E \left( e^{\displaystyle -q x^2(t)} \parallel y \right) =
 \int_{- \infty}^{+ \infty} {{d \xi} \over {\sqrt {2\pi}}} \  e^{\displaystyle
-{\xi^2 \over 2}} \
\ E \left( e^{\displaystyle -q 2D \xi^2 \tau_t} \parallel y \right)
\eqno(5.11)
$$

Using Eq (5.7) and the correspondance
 $\alpha=\displaystyle {{4D}}$ we get
$$
 E \left( e^{\displaystyle -q x^2(t)} \parallel y \right) =
\eqno(5.12)
$$
$$
{1 \over { \pi^2  {\sqrt {\pi  q}}}} \  { \left({1 \over y}\right)^{\mu +1}}
\int_0^{\infty} dk \ e^{ \displaystyle - {k^2 \over {4 q }}}
\int_{- \infty}^{+ \infty} d \nu \
e^{ \displaystyle - {  {Dt  } ( \mu^2 + \nu^2)}}  \nu \sinh \pi \nu \ \
 K_{i\nu} \left( k y  \right) K_{i\nu} ( k )
$$
For the free case $(\mu={1 \over 2})$, this result can readily be
recovered from the expression
of the Green's function $G_t (x,y)$ on the hyperbolic plane $^{(14)}$  through
 $$
 E \left( e^{\displaystyle -q x^2(t)} \parallel y \right) =
 \int_{-\infty}^{+\infty} dx \  \ e^{ \displaystyle -q x^2} \ {1 \over y^2} \
G_t (x,y)
\eqno(5.13)
$$
Let us finally mention that an alternative expression of the joint law
$\phi_t(\tau \parallel y) $ given in Eq (5.7) has been obtained by Yor
$^{(12)}$
through the time Laplace transform
$$
\int_0^{\infty} dt \ e^{-st} \ \phi_t(\tau \parallel y) ={1 \over {\tau
y^{1+\mu}}}
\ e^{- \left({{1+z^2}\over {\alpha \tau}}\right)} \ I_{\nu} \left({z \over
{\alpha \tau}}\right) \ \ \ \ \ \
\hbox{where} \ \ \nu=\sqrt{ \mu^2 +4 \displaystyle{s \over \alpha}}
\eqno(5.14)
$$
More generally, we refer to the mathematical literature  $^{(10)}$ $^{(11)}$
$^{(12)}$
 for different expressions related with probability distribution
of the functional $\tau\{T_s\}$ where $T_s$ is an independent time,
exponentially
distributed with parameter $s$.

\vskip 1 true cm

\leftline{ \bf 6. Conformal mapping from Poincar\'e half-plane to unit-disk}

\vskip 0.5 true cm

We introduce the conformal mapping from the Poincar\'e half upper-plane $\{
z=x+iy\ , y>0\}$
to the unit disk $\{w=re^{i\theta} , \vert r \vert \leq 1\}$
$$
 w= { {iz+1} \over {z+i}}
\eqno (6.1)
$$
The radial coordinate $r$ is directly related to the hyperbolic distance $d$ on
the Poincar\'e
half-plane between the arbitrary point $(x,y)$ and the point $\{x=0, y=1 \}$
that we choose as the initial point of Brownian
motion (3.9)
$$
r= \tanh \left({d \over 2} \right)
\eqno (6.2)
$$
The circle at infinity $r=1$ corresponds to the axis $y=0$.
The unit disk is very well suited to study the free hyperbolic Brownian motion
since it
contains explicitly the rotational invariance in the angle $\theta$.
Let us write in the new coordinates the metric
$$
 ds^2 = { {dx^2 +dy^2} \over y^2} =
\ { 4 \over {(1-r^2)^2}} \left( dr^2 + r^2 d\theta^2 \right)
\eqno (6.3)
$$
and the Laplace operator
$$
\Delta = y^2 \left( {\partial^2 \over {\partial x^2} } +{\partial^2 \over
{\partial y^2} }
\right) = \ { {(1-r^2)^2}  \over 4 } \left[ {1 \over r} {\partial \over
{\partial r} }
\left( r {\partial \over {\partial r}} \right) + {1 \over r^2}
 {\partial^2 \over {\partial \theta^2}}  \right]
\eqno (6.4)
$$

In the free case, the Fokker-Planck equation for the probability density
$Q_t(r,\theta)$
reads
$$
{ {\partial Q} \over {\partial t}} = { {\partial } \over {\partial r}} \left[
r { {\partial } \over {\partial r}} \left( { {(1-r^2)^2}  \over {4r} } Q
\right) \right]
+ { {(1-r^2)^2}  \over {4r^2} }{ {\partial^2 } \over {\partial \theta^2}} Q
  \eqno (6.5)
$$
The rules of stochastic calculus $^{(13)}$ give the corresponding
stochastic differential equations
$$  \hbox{ It\^o}   \left\{\matrix{
\displaystyle {dr\over dt}= D { {(1-r^2)^2}  \over {4r} }
+{ \sqrt {2D}} \ { {1-r^2} \over 2 } \ \eta_r(t)\hfill\cr
\noalign{\medskip}
\displaystyle {d\theta \over dt}={ \sqrt {2D}} \ { {1-r^2} \over {2r} } \
\eta_{\theta} (t)\hfill\cr
}\right\}
\eqno (6.6)
$$
$$ \hbox{Stratonovich} \left\{\matrix{
\displaystyle {dr\over dt}= D { {1-r^4}  \over {4r} }
+{ \sqrt {2D}} \ { {1-r^2} \over 2 } \ \eta_r(t)  \hfill\cr
\noalign{\medskip}
\displaystyle {d\theta \over dt}={ \sqrt {2D}} \ { {1-r^2} \over {2r} } \
\eta_{\theta} (t)\hfill\cr
}\right\}
\eqno (6.7)
$$
These equations cannot be integrated straightforwardly to give the processes
$\{r(t), \theta(t)\}$ as functionals of the white noises $\{\eta_r(t),
\eta_{\theta} (t)\}$
, unlike the representation (3.12) for the processes $\{x(t), y(t)\}$.
However, the asymptotic probability distribution $Q_{\infty}(r,\theta)$ in the
limit
$t \to \infty$ can be written directly from symmetry considerations. As
expected,
it is simply the uniform measure on the unit circle
$$
Q_{\infty}(r,\theta) = {1 \over {2 \pi}} \ \delta (r-1)
\eqno (6.8)
$$

For $\mu \neq {1 \over 2}$, the external constant drift $m=\left( \mu-{1 \over
2}\right)$
along direction $y$ breaks the rotational invariance in the angle $\theta$,
and the unit disk is not particularly well suited anymore. Nevertheless, it is
easy
to write the asymptotic probability distribution $Q_{\infty}(r,\theta)$ from
the
asymptotic law (4.12) for $P_{\infty}(x,y)$ and from the expression of the
Jacobian
$J\{(r,\theta) / (x,y)\}$
$$
Q_{\infty}(r,\theta) = {{4r} \over {(r^2+1-2r\sin \theta)^2}} \
P_{\infty}\left( { {2r\cos \theta} \over { r^2+1-2r\sin \theta }} ,
 {{1-r^2} \over { r^2+1-2r\sin \theta }} \right)
\eqno (6.9)
$$
We obtain after some algebra the generalisation of (6.8)
$$
Q_{\infty}(r,\theta) =\delta (r-1) \ {1 \over \sqrt{\pi} }
 { {\Gamma\left(\mu+{1 \over 2} \right)} \over \Gamma(\mu)} \ {1 \over
2^{\mu+{1 \over 2}}}
\ \big ( 1- \sin \theta \big)^{\mu-{1 \over 2}}
\eqno (6.10)
$$

\vskip 1 true cm

\leftline{ \bf 7. Conclusion}

\vskip 0.5 true cm

We have explored in great details the relations existing between,
on the one hand, the exponential functional $\tau_t$ that governs most of the
transport
properties of one dimensional classical diffusion
in a random potential $U(x)$, distributed as a Brownian motion  possibly with
some drift,
 and on the other, free or biased hyperbolic Brownian motion.
 As discussed in the introduction, this has to be considered as
 a special example of the more general link that connects
 one dimensional, or quasi-one dimensional, disordered
 systems, which usually admit some multiplicative stochastic stucture,
 and Brownian diffusion on symmetric spaces.

It is interesting to point out that the time relaxation spectrum found for the
probability distributions
$\psi_t(\tau)$ and $X_t(x)$ (Eqs 4.7  and 4.8) also appears for the quantum
spectrum of
a particle subject to a constant magnetic field $B$ on the hyperbolic plane
$^{(15)}$.
For the hyperbolic geometry, constant magnetic field is defined as the flux per
covariant surface element $ dS = \displaystyle {1 \over y^2} dx dy$.
The two spectra coincide if we choose the following correspondance between
magnetic field $B>0$ and drift $\mu$
$$
B  = \displaystyle {{1+\mu} \over 2}
$$
In this context, the existence of bound states in a strong enough magnetic
field
$B >  {1 \over{ 2 }}$ corresponds to the presence
of closed classical orbits $^{(15)}$.
It would be particularly interesting to understand more deeply this
correspondance
between spectra at the level of stochastic processes themselves.

As a final remark, let us mention that the
Liouville Hamiltonian that we encountered
in the path-integral formalism (5.3),
and that is closely related to hyperbolic geometry, also appears in the study
of refined properties of one dimensional quantum localisation for the
Schr\"odinger Hamiltonian
$H=- {d^2 \over dx^2} +V(x)$ where $V(x)$ is a Gaussian white noise potential
$^{(16)}$. Kolokolov uses a path-integral method to compute correlation
functions
of eigenstates and distribution function of inverse participation ratio in the
high energy limit.
In this formalism, Liouville Hamiltonian shows up in an effective action of
path-integral.
Expansion of this path-integral in a basis of eigenstates then gives
expressions very similar to the one we obtained for the probability
distribution of the joint
law of the processes $\{ \tau_t , U(t)\}$.

\vskip 1 true cm

\leftline{ \bf Acknowledgments}

\vskip 0.5 true cm

We are very grateful to Marc Yor, whose remark on the connection
between exponential functionals of Brownian motion of parameter $\mu = {1 \over
2}$
and free hyperbolic Brownian motion $^{(11)}$ is at the origin of this work. We
also
wish to thank him for having kindly given us his papers and other mathematical
references
related to this subject. We also thank Eugene Bogomolny for interesting
discussions.

\vskip 1 true cm

\leftline{ \bf Appendix : Path-integral method to prove the identity in law
(3.20) }

\vskip 0.5 true cm

In paragraph 3 we derived the identity in law (3.20) mentioned by Yor for the
free
hyperbolic Brownian motion $^{(11)}$ through a moments calculation. As
explained before,
this identity can easily be generalized
to any $\mu$ by the same method. Let us now derive it through a more
straightforward path-integral method using
an appropriate stochastic reparametrisation  of time in path-integral. This
tool has
already proven
to be very useful in other contexts $^{(17)}$.

 The integration of stochastic differential equations (3.27) gives the process
$x(t)$
 as the following functional of the two white noises $\{\eta_1, \eta_2\}$
 $$
  x(t) = { \sqrt {2D}} \int_0^t dv  \ \eta_1(v) \
e^{ \displaystyle - {2D \mu }v + {{ \sqrt {2D}} } \int_0^v \eta_2(s) ds }
 \eqno (A.1)
 $$
 The marginal law $X_t(x)$ of the process $x(t)$ therefore reads by definition
 $$
 X_t(x) = \int {\cal D} \eta_1(u) \ {\cal D} \eta_2(u) e^{\displaystyle -{1
\over 2} \int_0^t du \
 \big( \eta_1^2(u) +  \eta_2^2(u) \big)}
 $$
 $$
 \delta \bigg( x- { \sqrt {2D}} \int_0^t dv  \ \eta_1(v) \
  e^{ \displaystyle - {2D \mu }v + {{ \sqrt {2D}} } \int_0^v \eta_2(s) ds }
\bigg)
  \eqno (A.2)
  $$
 Let us first change from $\eta_2(u)$ to $U(u)$ defined by
 ${\beta \over 2}  U(v) = - {2D \mu }v + {{ \sqrt {2D}} } \int_0^v \eta_2(s)
ds$
 $$
  X_t(x) = \int {\cal D} \eta_1(u) \  e^{\displaystyle -{1 \over 2} \int_0^t du
\  \eta_1^2(u)}
  \int_{U(0)=0} {\cal D} U(u) e^{\displaystyle -{1 \over 4D} \int_0^t du \
  \left( { {\beta \over 2}  { dU \over dv } + 2D \mu } \right)^2 }
$$
$$
\delta \bigg( x- { \sqrt {2D}} \int_0^t dv  \ \eta_1(v) \
  e^{ \displaystyle  {\beta \over 2}  U(v)    } \bigg)
\eqno (A.3)
 $$
And now from $\eta_1(u)$ to $x(u) \equiv { \sqrt {2D}} \int_0^u dv  \ \eta_1(v)
\
  e^{ {\beta \over 2}  U(v) }$
  $$
    X_t(x) = \int_{x(0)=0}^{x(t)=x} {\cal D} x(u) \int_{U(0)=0} {\cal D} U(u) \
    e^{\displaystyle -{1 \over 4D} \int_0^t du \  \left( {
  {\beta \over 2}  { dU \over dv } + 2D \mu} \right)^2}
   $$
  $$
  e^{\displaystyle -{1 \over 4D} \int_0^t du \  \left( {dx \over du}
e^{\displaystyle
  -{\beta \over 2} U(u)} \right)^2}
   \eqno (A.4)
   $$
Let us now perform a time-reparametrisation of the trajectories $x(u)$ in order
to recover
the Wiener measure
$$
\int_0^t du \ \left( {dx \over du} \right)^2 e^{-{\beta} U(u)}= \int_0^{\tau}
ds \
\left( {dx \over ds} \right)^2
 $$
where
$$
 Jds =e^{\beta U(u)} du \ \ \ \ \hbox{and} \ \ \ \tau\{U(u)\} = \int_0^t
e^{\beta U(u)} du
 \eqno (A.5)
 $$
The new final time $\tau \{U(u)\}$ is not fixed anymore, but depends on the
realization of
random potential $U(u)$.
To take into account this constraint, we can insert the identity
$$
1= \int_0^{\infty} d\tau \ \delta \left( \tau - \int_0^t e^{\beta U(u)} du
\right)
 \eqno (A.6)
 $$
to obtain
$$
X_t(x) = \int_0^{\infty} d\tau  \int_{x(0)=0}^{x(t)=x} {\cal D} x(u)
\int_{U(0)=0} {\cal D} U(u) \
    e^{\displaystyle -{1 \over 4D} \int_0^t du \  \left( {
  {\beta \over 2}  { dU \over dv } + 2D \mu} \right)^2}
  $$
 $$
  e^{\displaystyle -{1 \over 4D} \int_0^\tau ds \  \left( {dx \over ds}
   \right)^2} \delta \left( \tau - \int_0^t e^{\beta U(u)} du \right)
  \eqno (A.7)
  $$
Let us now perform the Gaussian path-integral on $x(u)$
$$
\int_{x(0)=0}^{x(t)=x} {\cal D} x(u) e^{\displaystyle -{1 \over 4D} \int_0^\tau
ds \  \left( {dx \over ds}
   \right)^2} = {1 \over { \sqrt{ 4 \pi D \tau}}} \  e^{\displaystyle -{x^2
\over {4Dt}}}
 \eqno (A.8)
 $$
to get
$$
X_t(x) = \int_0^{\infty} d\tau {1 \over { \sqrt{ 4 \pi D \tau}}}  \
e^{\displaystyle
-{x^2 \over {4Dt}}} \ \psi_t(\tau)
 \eqno (A.9)
 $$
where
$$
\psi_t(\tau) = \int_{U(0)=0} {\cal D} U(u)
    e^{\displaystyle -{1 \over 4D} \int_0^t du \  \left( {
  {\beta \over 2}  { dU \over dv } + 2D \mu} \right)^2}
\delta \left( \tau - \int_0^t e^{\beta U(u)} du \right)
 \eqno (A.10)
 $$
is by definition the probability distribution of the functional
$\tau_t = \int_0^t e^{\beta U(u)} du$. \hfill \break
Eq (A.9) is just the translation in terms of probability distributions of the
identity in law
between the process $x(t)$ and a linear Brownian motion of stochastic time
$\tau_t$
$$
x(t) ={ \sqrt {2D}} \int_0^{\displaystyle \tau_t} du \ \eta(u)
$$

\vskip 1 true cm

\leftline{ \bf References }

\vskip 0.5 true cm

\item{\hbox to\parindent{\enskip 1. \hfill}}
M.E. Gertsenshtein and V.B. Vasil'ev, {\it Theor. Prob. Appl.} {\bf 4} (1959)
391;
F.I. Karpelevich, V.N. Tutubalin and M.G. Shur, {\it Theor. Prob. Appl.} {\bf
4} (1959) 399;
G.C. Papanicolaou, {\it SIAM J. Appl. Math.} {\bf 21} (1971) 13.

\item{\hbox to\parindent{\enskip 2. \hfill}}
O.N. Dorokhov, {\it JETP Lett.} {\bf 36} (1982) 318;
O.N. Dorokhov, {\it Sov. Phys. JETP} {\bf 58} (1983) 606;
P.A. Mello, P. Pereyra and N. Kumar, {\it Ann. Phys.} {\bf 181} (1988) 290.

\item{\hbox to\parindent{\enskip 3. \hfill}}
A. H\"uffmann, {\it J. Phys. A : Math. Gen. } {\bf 23} (1990) 5733;
C.W.J.Beenaker and B.Rejaei, {\it Phys.Rev.Lett.} {\bf 71} (1993) 3689.

\item{\hbox to\parindent{\enskip 4. \hfill}}
N. Balazs and A. Voros, {\it Phys. Rep.} {\bf  143} (1986) 109;
M.C. Gutzwiller, ``Chaos in classical and quantum mechanics", Springer Verlag
(1990).

\item{\hbox to\parindent{\enskip 5. \hfill}}
A.A. Chernov, {\it Biophysics} {\bf 12} (1967) 336;
F. Solomon, {\it Ann. Prob.} {\bf 3} (1975) 1;
H. Kesten, M. Koslov and F. Spitzer, {\it Compositio Math. } {\bf 30 } (1975)
145;
Y.A.G. Sina\"{\i}, {\it Theor. Prob. Appl.} {\bf XXVII} (1982) 256;
B. Derrida and Y. Pomeau,{\it Phys. Rev. Lett. } {\bf 48 } (1982) 627;
B. Derrida, {\it J. Stat. Phys.} {\bf 31} (1983) 433;
K. Kawazu and H. Tanaka, {\it Sem. Prob.} {\bf XXVII} (1993).

\item{\hbox to\parindent{\enskip 6. \hfill}}
B. Derrida and H.J. Hilhorst, {\it J. Phys. A } {\bf  16 } (1983) 2641;
C. de Callan, J.M. Luck, Th. Nieuwenhuizen and D. Petritis,
{\it J. Phys. A} {\bf  18 } (1985) 501;
J.M. Luck, ``Syst\`emes d\'esordonn\'es unidimensionnels", Collection Al\'ea
Saclay (1992).

\item{\hbox to\parindent{\enskip 7. \hfill}}
J.P. Bouchaud and A. Georges, {\it Phys. Rep.} {\bf 195} (1990) 127;
J.P. Bouchaud, A. Comtet, A. Georges and P. Le Doussal,
 {\it Ann. Phys. } {\bf 201} (1990) 285;
A. Georges, th\`ese d'\'etat de l'Universit\'e Paris 11 (1988).

\item{\hbox to\parindent{\enskip 8. \hfill}}
S.F. Burlatsky, G.H. Oshanin, A.V. Mogutov and M. Moreau,
 {\it Phys. Rev. A} {\bf 45 } (1992) 6955;
G. Oshanin, A. Mogutov and M. Moreau, {\it J. Stat. Phys.} {\bf 73 } (1993);
G. Oshanin, S.F. Burlatsky, M. Moreau and B. Gaveau, {\it Chem. Phys.} {\bf 178
} (1993).

\item{\hbox to\parindent{\enskip 9. \hfill}}
C. Monthus and A. Comtet, {\it J. Phys. I (France)} {\bf 4} (1994), 635.

\item{\hbox to\parindent{\enskip 10. \hfill}}
D. Dufresne,  {\it Scand. Act. J. } {\bf } (1990) 39;
A. de Schepper, M. Goovaerts and F. Delbaen,
 {\it Ins.: Math. and Eco.} {\bf 11} (1992) 291;
H. Geman and M. Yor, {\it Math. Fin.} {\bf 3} (1993) 349.

\item{\hbox to\parindent{\enskip 11. \hfill}}
M. Yor,  {\it Adv. Appl. Prob.} {\bf 24} (1992) 509.

\item{\hbox to\parindent{\enskip 12. \hfill}}
M. Yor, {\it Ins. Math. Econ.} {\bf 13} (1993).

\item{\hbox to\parindent{\enskip 13. \hfill}}
C.W. Gardiner, ``Handbook of stochastic methods for physics, chemistry and the
natural sciences"
(Second edition), Springer Verlag (1990).

\item{\hbox to\parindent{\enskip 14. \hfill}}
A. Terras, ``Harmonic analysis on symmetric spaces and Applications I",
Springer Verlag (1985).

\item{\hbox to\parindent{\enskip 15. \hfill}}
A. Comtet,  {\it Ann. Phys.} {\bf 173 } (1987) 185;
J.E.Avron and A.Pnueli, in  ``Ideas and methods in Quantum and Statistical
Physics", vol. 2,
Cambridge University Press (1992).

\item{\hbox to\parindent{\enskip 16. \hfill}}
I. Kolokolov, {\it Sov. Phys. JETP} {\bf 76} (1993) 1099;
I. Kolokolov, {\it Europhys. Lett.} {\bf 28}

\item{\hbox to\parindent{\enskip 17. \hfill}}
I.H. Duru and H. Kleinert,  {\it Phys. Lett.} {\bf 84B} (1979) 185;
P. Blanchard and M. Sirugue,  {\it J. Math. Phys.} {\bf 22} (1981) 1372;
I.H. Duru, {\it Phys. Rev. D} {\bf 28} (1983) 2689;
A. Young and C. De Witt-Morette, {\it Ann. Phys.} {\bf 169} (1986) 140;
D.C. Khandekar, S. V. Lawande and K.V. Bhagwat, ``Path integral methods and
their applications",
World scientific, Singapore (1993).

\end